\documentstyle[floats,aps,prl,twocolumn]{revtex}

\begin{document}
\draft
\wideabs{
\title{On Effect of Equilibrium Fluctuations on Superfluid Density in
Layered Superconductors}
\author{S.N. Artemenko and S.V.Remizov}
\address{
Institute for Radioengineering and Electronics of Russian Academy of
Sciences, Mokhovaya 11, 103907 Moscow, Russia }

\date{\today}
\maketitle
\begin{abstract}
We calculate suppression of inter- and intralayer superconducting currents
due to equilibrium phase fluctuations and find that, in contrast to a recent
prediction, the effect of thermal fluctuations cannot account for linear
temperature dependence of the superfluid density in high-T$_c$
superconductors at low temperatures. Quantum fluctuations are found to
dominate over thermal fluctuations at low temperatures due to hardening of
their spectrum caused by the Josephson plasma resonance. Near T$_c$ sizeable
thermal fluctuations are found to suppress the critical current in the stack
direction stronger, than in the direction along the layers. Fluctuations of
quasiparticle branch imbalance make the spectral density of voltage
fluctuations at small frequencies non zero, in contrast to what may be
expected from a naive interpretation of Nyquist formula.
\end{abstract}
}

One of important problems in layered high-T$_c$ cuprate superconductors that
is still under discussion is the origin of the observed linear
low-temperature dependence of the superfluid density (phase stiffness). The
latter is directly related to magnetic penetration depths, $\lambda_\|$ and
$\lambda_\perp$, for a magnetic field screened by currents flowing in
directions parallel and perpendicular to the superconducting planes,
respectively~\cite{exp}. This dependence is usually attributed to
contribution of quasiparticles near the nodes of the $d$-wave gap. According
to the alternative explanation suggested in refs.~\cite{RS,EK} the linear
decrease in the temperature dependence of $1/\lambda^2$ is induced entirely
by classical thermal phase fluctuations. Recently the role of fluctuations
was reconsidered~\cite{PRRM} for a $d$-wave superconductor by means of a
microscopic approach within a functional integral framework, and quantum
phase fluctuations were found to lead to a sizeable renormalization of the
superfluid density, the effect of thermal fluctuations being small for $T <
T_c$.

Thermal and quantum fluctuations considered in refs.~\cite{RS,EK,PRRM} are
equilibrium fluctuations, therefore, the problem can be solved by means of a
phenomenological approach based on the fluctuation--dissipation theorem.
Such an approach is more simple and physically transparent, being in the
same time more general because it is not restricted to a specific model of
high-T$_c$ superconductivity. In the present work we reconsider the role of
equilibrium phase fluctuations in layered superconductors applying the
fluctuation-dissipation theorem to equations of the linear response, and
come to conclusions different from those of refs.~\cite{RS,EK,PRRM}. Our
calculations demonstrate that, in agreement with the results of
ref.~\cite{PRRM}, at low temperatures quantum fluctuations dominate over
thermal fluctuations due to Coulomb effects. This happens because
non-uniform phase fluctuations induce a fluctuating electric field resulting
in a finite energy of such fluctuations which is reflected in a finite value
of the Josephson plasma frequency. So unlike refs.~\cite{RS,EK} we assert
that a contribution of thermal fluctuations cannot account for temperature
dependence of the penetration depth. However, in contrast to the results of
ref.~\cite{PRRM} we find that renormalization of the superfluid density due
to fluctuations at low temperatures is not large. Near T$_c$, when
quasiparticle density dominates over the superfluid density, we find a
different picture. In this case thermally induced fluctuations of the
quasiparticle branch imbalance are found to be important, and a reduction of
the superfluid density may become sizeable. We find that a suppression of
the superconducting critical current due to thermal fluctuations is larger
in the stack direction, than in the direction parallel to the conducting
layers. This points out a possibility of a destruction of the
superconductivity in the perpendicular direction at temperatures lower, than
the critical temperature for the in-plane direction.

The fluctuation-dissipation theorem relates correlation functions of
fluctuations to dissipative parts of kinetic coefficients describing the
linear response of a system. So we start with the linear response of a
layered superconductor. We consider expressions for current and charge
density written in a form of functions of gauge invariant vector and scalar
potentials. These potentials are
\begin{equation}
  {\bf P}_s
  =
  (1/2)\nabla \chi - (1/c){\bf A},
  \quad
  \mu =(1/2) \partial_t \chi + \Phi
,\label{pmu}
\end{equation}
where ${\bf A}$ is the vector potential, $\chi$ is the phase of the order
parameter, and $\Phi$ is the electric potential (we use units with $e=1$,
$\hbar=1$, and $k_B=1$). The gauge invariant vector potential plays a role
of the superconducting momentum, while the gauge invariant scalar potential
is responsible for branch imbalance~\cite{T} and charging effects. The
electric field is expressed in terms of the potentials as
\begin{equation}
  {\bf E}
  =
  \partial_t {\bf P}_s - \nabla \mu
. \label{E}
\end{equation}

In layered superconductors the variables above are related to specific
layers and must be supplied by layer numbers, {\it e. g.}, the parallel
component of ${\bf P}_s$ will be denoted as ${\bf P}_n$. Furthermore, the
role of the perpendicular component of ${\bf P}_s$ plays the gauge invariant
phase difference between the layers, $\varphi_{n} = \chi_{n+1} - \chi_{n} -
2 s A_\perp/c$, where $s$ is the lattice period in the stack direction. So,
we may denote $P_{\perp n} = \varphi_{n}/2s$.

Equations determining current and charge densities read
\begin{eqnarray}
  &&
  {\bf j}_{\| n}
  =
  \frac{c^2}{4\pi \lambda_\|^2}{\bf P}_n
  +
  \sigma_{0\|} \partial_t {\bf P}_n
  -
  \sigma_{1\|} \nabla_\| \mu_n
, \label{jl}\\
  &&
  j_{\perp n}
  =
  j_c \sin{\varphi_n}
  +
  \frac{\sigma_{0\perp}}{2 s} \partial_t \varphi_n
  -
  \sigma_{1\perp}\nabla_n
  \mu_n
, \label{jt} \\
  &&
  \partial_t \rho_n
  =
  (\gamma \partial_t + \nu_b)
  \frac{\kappa^2}{4\pi} \mu_n
  -
  (\sigma_{2\|}\nabla_\|^2
  +
  \sigma_{2\perp}\nabla_n^2)\mu_n
  +
\nonumber \\
  &&
  \partial_t(\sigma_{1\|}\nabla_\|{\bf P}_n
  +
  \sigma_{1\perp} \nabla_n
  \varphi_n/2s)
, \label{mu}
\end{eqnarray}
where $\nabla_n \mu_n = (\mu_{n+1} - \mu_n)/s$ is the discrete version of
the spatial derivative in the transverse direction, $\kappa^{-1}$ is the
Thomas--Fermi screening radius, and $\nu_b$ is the branch imbalance
relaxation rate.

The first terms in equations (\ref{jl}-\ref{jt}) describe the
superconducting current for relative directions, and the rest terms are
related for quasiparticle contributions. Equation (\ref{mu}) for charge
density can be interpreted as a continuity equation for quasiparticles. It
was derived for conventional superconductors in ref.~\cite{AV} and
generalized later for layered superconductors with $s$-wave~\cite{AKpl} and
$d$-wave~\cite{AKd} pairing.

Note that the quasiparticle current does not have the form $j=\sigma E$
valid for a normal state. Instead, the response of a superconductor to the
components of the field $E$ expressed in terms of the temporal derivative of
${\bf P}_s$ and of the spatial derivative of $\mu$ ({\it cf.} (\ref{E})) are
described by different generalized conductivities $\sigma_{i}, i=0,1,2$. The
respective parts of the electric field $E$ are induces by time dependent
perturbations of current density, and by perturbations of charge density.

Explicit expressions for the coefficients in equations (\ref{jl}-\ref{mu})
depend on temperature, on scattering, and on a mechanism of the
superconductivity (see \cite{AKpl,AKd}, note that the explicit expressions
for coefficients in ref.~\cite{AKd} were presented in the dynamic limit
$\omega > Dq^2$ where $D$ is a diffusion coefficient in quasiclassic
approximation). At temperatures $T \rightarrow 0$ all conductivities are
small ({\it e. g.} in a $s$-wave superconductor they are exponentially
small), and $\gamma \rightarrow 1$. At $T \rightarrow T_c$ the
conductivities $\sigma_i$ approach the normal state conductivities.
Furthermore, the differences between $\sigma_i$ for a given direction
vanishes as $\Delta/T$, $\gamma \rightarrow \Delta/T, \, \nu_b \rightarrow
\Delta/T$, so that equation (\ref{mu}) approaches the continuity equation as
$\Delta \rightarrow 0$. Strictly speaking products in (\ref{jl}-\ref{mu})
assume the convolution with respect to time and coordinates, {\it i. e.} in
the Fourier transformed form the coefficients in the equations are frequency
and wave vector dependent.

In order to derive equations relating fluctuations $\delta\mu$,
$\delta\varphi$ and $\delta{\bf P}$ with fluctuations of charge and current
densities, $\delta \rho$ and $\delta {\bf j}$, we insert expression
(\ref{jl}-\ref{mu}) into Maxwell equation $$\nabla \times {\bf H}= \frac{4
\pi}{c}{\bf j}+ \frac{1}{c}\partial_t {\bf D}$$ and into the Poisson
equation. Then making Fourier transformation with respect to time, to
in-layer coordinates ${\bf r}_\|$, and to layer numbers $n$, we find
\begin{eqnarray}
  &&
  \left(
    \begin{array}{c}
      \delta j_\perp
      \\
      \delta j_\|
      \\
      \delta \rho
    \end{array}
  \right)
  =
  \hat A
  \left(
    \begin{array}{c}
      \delta \varphi /2s
      \\
      \delta {\bf P}_\|
      \\
      \delta \mu
    \end{array}
  \right)
, \label{main} \\
  &&
  \hat A \!
  =
  \! \frac{4}{\pi}\!
  \left(\!
    \begin{array}{ccc}
      \epsilon
      (\omega \omega_{0}\! - \!\tilde{\omega}_p^2)
      &
      c^2 q_\| \hat q_\perp
      &
      \epsilon \hat q_\perp \omega_{1}
      \\
      c^2 q_\| \hat q_\perp
      &
      \omega \Omega_{0}\! -\! \tilde{\Omega}_p^2
      &
      q_\| \Omega_{1}
      \\
      \epsilon \hat q_\perp \omega_{1}
      &
      q_\| \Omega_{1}
      &
      (\gamma \!- \!\frac{\nu_b}{i\omega}) \kappa^2\!
      +
      \! \frac{\epsilon \hat q_\perp^2 \omega_{2} + q_\|^2 \Omega_{2}}{\omega}
    \end{array} \!
  \right)
. \nonumber
\end{eqnarray}
Here $\hat q_\perp = (2/s)\sin{(q_\perp s/2)}$, where $|q_\perp|<\pi/s$ is
the wave number obtained from the discrete Fourier transformation with
respect to layer numbers, $\omega_{i} = \omega +i \omega_{ir}$, $\Omega_{i}
= \omega +i \Omega_{ir}$, $\omega_{ir}=4\pi\sigma_{i\perp}\epsilon$ and
$\Omega_{ir}= 4\pi\sigma_{i\|}$ are dielectric relaxation frequencies.
Furthermore, $\tilde{\omega}_p^2=\omega_p^2 (1+\lambda_\perp^2q_\|^2)$,
$\tilde{\Omega}_p^2=\Omega_p^2 (1+\lambda_\|^2 \hat q_\perp ^2)$, where
$\Omega_p =c/\lambda$ and $\omega_p= c/\lambda_\perp\sqrt{\epsilon}$ are the
plasma frequencies for directions parallel and perpendicular to the layers,
$\epsilon$ is a dielectric constant in transverse direction, $\lambda_\perp
=c/\sqrt{8\pi s j_c}$. The in-layer plasma frequency $\Omega_p$, is much
larger than typical frequencies of the problem which are of order of the
Josephson plasma frequency, $\omega_p$.

Note that matrix $\hat A$ satisfy the Onsager symmetry relations. Now we
apply the fluctuation-dissipation theorem ({\it cf.} \cite{LL}) to equations
(\ref{main}).

From the expression for energy dissipation density $Q$ which in
superconductors is given by \cite{AG}
\begin{equation}
  Q
  =
  \rho \partial_t \mu + {\bf j}\partial_t {\bf P}_s
, \label{sc}
\end{equation}
one can identify potentials $\mu$ and ${\bf P}_s$ with generalized forces
related to variables $\rho$ and ${\bf j}$, respectively. So according to the
fluctuation-dissipation theorem correlation functions of $\delta j_\perp$,
$\delta j_\|$ and $\delta \rho$ are determined by imaginary parts of related
coefficients of the matrix $\hat A$ in (\ref{main}). For example, we find
\begin{equation}
  \langle \delta j_\perp ({\bf q}, \omega) \delta j_\perp ({\bf q}', \omega') \rangle
  =
  (2 \pi)^4 (j_\perp^2)_\omega \delta (\omega + \omega') \delta ({\bf q} + {\bf q}')
\label{cor}
\end{equation}
with $(j_\perp^2)_\omega =2\tilde T \sigma'_{0\perp}/s$, where $\tilde T =
(\omega/2) \coth{(\omega/2T)}$, $\sigma' =\Re{\sigma}$. Similarly, $(j_\perp
\rho)_\omega =2\tilde T \hat q_\perp \sigma'_{1\perp} /s \omega$,
$(\rho^2)_\omega =2\tilde T (\nu_b/4\pi + q_\perp^2 \sigma'_{2\perp} +
q_\|^2 \sigma'_{2\|})/s \omega^2$ and so on. An alternative way to calculate
correlation functions based on the Langevin sources in equations of motion,
and on the dissipation function of the system gives similar results.

Correlation functions of $\delta{\bf P}_s$ and $\delta\mu$ are related to
the inverse matrix $\hat A^{-1}$. For example, using the upper diagonal
component of $\hat A^{-1}$ we can find the spectral density of phase
difference fluctuations
\begin{equation}
  (\delta \varphi^2)_\omega
  =
  \Im \frac{8\pi \tilde T(\omega \Omega_{0}\!
  -\!
  \tilde{\Omega}_p^2)[(\gamma -\frac{\nu_b}{i\omega}) \kappa^2\!
  +\!
  \frac{\epsilon \hat q_\perp^2 \omega_{2} + q_\|^2 \Omega_{2}}{\omega} ]}{D}
  \label{fio}
\end{equation}
where $D$ is the determinant of the matrix $\hat A$.

Zeros of $D$ determine collective modes and penetration of electric and
magnetic fields into a superconductor. They are important in calculation of
spectral density of voltage noise. The voltage between contacts separated by
$N$ layers,  can be expressed according to eq.(\ref{E}) as $V =
\partial_t \sum\limits_{n=0}^{N-1} \varphi_n/2 -(\mu_N -\mu_0)$. Then
we make Fourier transformation and calculate mean square value of the voltage
using (\ref{fio}) and similar expressions for other correlation functions.
We assume that the width of the contacts is large enough, therefore, we need
functions at $q_\|=0$. Finally, after some algebra we find the expression
for the spectral density of voltage fluctuations
\begin{eqnarray}
  &&
  (\delta V^2)_\omega
  =
  \frac{16 \tilde T}{S\varepsilon}\int dq_\perp
  \frac{\sin^2{\frac{q_\perp L}{2}}}{\hat q_\perp^2} \times
\nonumber
  \\ &&
  \Im
  \frac
  {
    \omega(\omega \gamma + i\nu_b)\kappa^2
    +
    \epsilon \hat q_\perp^2[\omega(\omega_0+\omega_2- 2\omega_1)-\omega_p^2]
  }
  {
    (\omega_p^2
    \!-\!
    \omega \omega_0)(\omega \gamma
    \!+\!
    i\nu_b)\kappa^2
    \! +\!
    \epsilon \hat q_\perp^2 [\omega(\omega_1^2
    \!-\!
    \omega_0 \omega_2)
    \!+\!
    \omega_2\omega_p^2]
  }
\label{vo}
\end{eqnarray}
with $L=Ns$.

At low temperatures, $T \ll \Delta$, the relaxation frequencies are small in
comparison to plasma frequencies $\omega_p$, and zero of the denominator
gives the underdamped Josephson plasma mode \cite{AKpl,plth,plex} at
$q_\|=0$. In the opposite limit $T \gg \Delta$, near T$_c$, the plasma
frequencies are small, $\Omega_p^2, \omega_p^2 \propto N_s \propto
(\Delta/T)^2$ where $N_s$ is a fraction of the condensed electrons. On the
other hand, all conductivities $\sigma_{i\|}$ and $\sigma_{i\perp}$ at $T
\rightarrow T_c$ approach the normal state conductivities for parallel and
perpendicular directions, respectively. So near T$_c$ the dielectric
relaxation frequencies are larger, than plasma frequencies for respective
directions. Furthermore, calculations for $s$- and $d$-wave pairing give
$\gamma =\pi \Delta/4T$ \cite{AV} and $\gamma =\pi
\Delta_0/2T\sqrt{i\nu/\omega}$ \cite{AKd}, respectively. The branch
imbalance relaxation rate is determined by energy relaxation for $s$-wave
pairing and by elastic scattering for $d$-wave pairing, and in both cases
$\nu_b \propto \Delta/T$. Then zeros of the denominator in equation
(\ref{vo}) give the Carlson-Goldman mode~\cite{CG} for direction
perpendicular to the layers, that is underdamped in the case of isotropic
pairing in a narrow frequency region, $\nu_b,\omega_p^2/\omega_r \ll \omega
\ll (T/\Delta)\omega_p^2/\omega_r$. The spectrum of the mode is $\omega^2
\approx \epsilon q_\perp^2 \gamma/\kappa^2$. This spectrum in isotropic
superconductors was found in clean limit in ref.~\cite{AVco} and in dirty
limit in ref.~\cite{SS}. Evolution of the spectrum from the Josephson plasma
mode to the anisotropic Carlson--Goldman mode in layered superconductors was
studied in ref.~\cite{AKpl}. In the case of $d$-wave pairing $\gamma$ is not
real, and the Carlson--Goldman mode is never underdamped ({\it cf.}
\cite{AKd}).

For $N \gg 1$ the leading contribution to the voltage fluctuations is
\begin{equation}
  (\delta V^2)_\omega^{(0)}
  =
  \frac{2 \tilde T L}{S \sigma_{0\perp}}
  \frac
    {\omega^2 \omega_{0r}^2}
    {(\omega_p^2-\omega^2)^2 + \omega^2 \omega_{0r}^2}.
\label{vo1}
\end{equation}
This expression corresponds to the Nyquist formula and exhibits the
Josephson plasma resonance. At zero frequency this contribution vanishes,
but there is an additional contribution to voltage fluctuations which
remains finite at low frequencies $\omega\ll \nu_b$. The latter contribution
is related to the quasiparticle branch imbalance fluctuations, and is
especially pronounced near $T_c$:
\begin{equation}
  (\delta V^2)_\omega^{(1)}
  =
  \frac{4 T }{S \sigma_\perp}
  \frac{l_E^2}{\sqrt{4 l_E^2 + s^2}}
, \label{vo2}
\end{equation}
where $l_E^2 =4\pi \sigma_\perp/\nu_b \kappa^2$ (we assumed a thick sample,
$L \gg l_E$). Equation (\ref{vo2}) contradicts to a naive interpretation of
the Nyquist theorem, according to which voltage noise at zero frequency is
absent because the static resistivity of a superconductor is equal to zero.
The non-zero contribution is related to a voltage drop near the
superconductor boundary due to penetration of the electric field into
layered superconductor over a distance needed for the branch imbalance
relaxation. Though $(\delta V^2)_\omega$ in (\ref{vo2}) is independent on
the distance $L$ between the contacts, it may give a sizeable contribution
which can be measured easily in small mesa structures.

Using equation (\ref{fio}) we can calculate mean square fluctuation of the
phase difference.
\begin{equation}
  \langle \delta \varphi^2 \rangle
  =
  \int (\delta \varphi^2)_\omega \frac{d\omega d{\bf q}}{(2\pi)^4}
. \label{fi}
\end{equation}
Since $\langle j_c \sin{(\varphi + \delta \varphi)} \rangle \approx j_c
(1-\langle \delta \varphi^2 \rangle/2) \sin{\varphi}$, the latter term
determines renormalization of the superfluid density. We calculate this
renormalization, first, in the limit of low temperature.

At $T \ll \Delta$ the relaxation frequencies are small in comparison to the
related plasma frequencies $\omega_p$ and $\Omega_p$. We assume a
simplifying condition $\kappa s/\epsilon \gg 1$ which holds in high-T$_c$
superconductors, therefore, perturbations of $\delta \mu$ are small and can
be neglected. Then the leading terms in the denominator of the matrix $\hat
A$ are $$D=\omega \kappa^2 \Omega_p^2 \{\epsilon
(\omega_p^2-\omega^2)(1+\lambda_\|^2 \hat q_\perp ^2)+ c^2q_\|^2$$ $$ -4\pi
i\omega[\sigma_{0\perp}(1+\lambda_\|^2 \hat q_\perp ^2) +
\sigma_{0\|}\epsilon (\omega_p^2-\omega^2
+c^2q_\|^2\epsilon)/\Omega_p^2]\}.$$ A phase volume near zeros of $D$
related to the Josephson plasma mode contributes much to the integral
(\ref{fi}) resulting in a finite fluctuations at $T=0$. Furthermore, the
integral over $q_\|$ diverges at large $q_\|$, and we cut-off it at $q_\|
\sim 1/\xi_0$, where $\xi_0$ is the superconducting correlation length. In
dimensional units we obtain
\begin{eqnarray}
&&
  \langle \delta \varphi^2 \rangle
  \approx
  \frac{16}{\pi \sqrt{\epsilon}}\frac{e^2}{\hbar c}
  \frac{\lambda_\|}{\xi_0}
  \left(
    1 + \frac{T^2}{T_0^2}
  \right)
, \label{fif} \\
  &&
  T_0
  \approx
  \frac{\hbar\Omega_p}{2\pi k_B}
  \sqrt{
    \frac
      {3 \Omega_p s}
      {2\sigma_{0\|}\xi_0\sqrt{\epsilon}\ln{\lambda_J/\xi_0}}
  }
. \nonumber
\end{eqnarray}
This result differs from that obtained in refs.~\cite{RS,EK,PRRM}. For
parameter values typical for layered cuprates we find $T_0 \gg T_c$. The
magnitude of $\langle \delta \varphi^2 \rangle$ is not large because the
large factor $\lambda/\xi_0$ in (\ref{fif}) is multiplied by the small fine
structure constant. Thus the renormalization of the penetration depths due
to phase fluctuations at low temperatures is practically temperature
independent, and is not large.

Calculation of $\langle \delta P_\|\rangle ^2 \xi_0^2 $ which determines
suppression of the superfluid density in the in-plane direction gives value
similar to (\ref{fif}).

Now we calculate fluctuations at high temperatures, $T \to T_c$. Since at
such temperatures fluctuations are quasi-stationary they can be found using
the standard approach based on functional integration of the free energy of
the system, which includes energy of the magnetic field and of the
superconducting current. We calculate $\langle \delta \varphi^2 \rangle$ and
$\langle \delta P_\|^2 \rangle \xi_0^2$, which determine the renormalization
of the superfluid stiffness in perpendicular and parallel directions,
respectively, using again the cut-off at $q_\| \sim 1/\xi_0$. Then in the
limit $\xi_0 \ll \lambda_J$, which definitely holds for layered high-T$_c$
superconductors, we find in dimensional units
\begin{equation}
  \langle \delta \varphi^2 \rangle
  \approx
  \frac {16 e^2 k_B T \lambda_\|^2}{\hbar^2 c^2 s}
  \ln{\frac{\lambda_\perp}{\xi_0}}
  \approx
  4 \ln{\frac{\lambda_\perp}{\xi_0}}\, \langle \delta P_\|^2 \rangle \xi_0^2
. \label{flh}
\end{equation}
Since in layered high-T$_c$ superconductors $\lambda_\perp$ is by few orders
of magnitude larger, than the correlation length $\xi_0$, the logarithm in
equation (\ref{flh}) is large as well. Then suppression of the critical
current by fluctuations in the direction perpendicular to the layers is
larger, than in the parallel direction. Using parameters typical for BSCCO,
$\lambda_\|(0) \approx 1.5 \times 10^{-5}$ cm, $s \approx 1.5 \times
10^{-7}$ cm and $\lambda_\perp/\xi_0 \approx 10^4$, we estimate $$ \langle
\delta \varphi^2 \rangle \approx 0.5 \left[ \frac{\lambda_\|(T)}
{\lambda_\|(0)}\right]^2 \frac{T}{80\mbox{K}}. $$ Since
$[\lambda_\|(T)/\lambda_\|(0)]^2$ diverges as $T \rightarrow T_c$ we
conclude that thermal fluctuations may lead to a sizeable reduction of the
critical current. Note that we study fluctuations in the linear
approximation and, hence, do not take into account that the phase
perturbations due to thermal fluctuations may overcome a finite potential
barrier. The latter process would result in a destruction of a
superconducting current analogous to Josephson junctions \cite{flJ}.

Observation of smaller values of T$_c$ in the stack direction, than in
direction parallel to the layers, was reported in YBCO single crystals with
low oxygen content \cite{Zv}. Such crystals are expected to have large
$\lambda_\|$, which according to (\ref{flh}) is in favor of large
fluctuations. However, it is difficult to explain by the fluctuation
mechanism so large differences between critical temperatures observed in
ref.~\cite{Zv} for different directions.

We are grateful to A. A. Varlamov for attracting our attention to the
results of ref.~\cite{EK} and for useful discussion, and to V. F.
Gantmakher, K. E. Nagaev and A. Ya. Shulman for useful discussion. This work
was supported by project 98-02-17221 of Russian Foundation for Basic
Research, and by project 96053 of Russian program on superconductivity.

\end{document}